\documentclass[aps,prb,twocolumn,superscriptaddress]{revtex4}

\usepackage[dvips]{graphicx}
\usepackage[dvips]{color}
\usepackage{amsmath}

\renewcommand{\Vec}[1]{{\bf #1}}

\begin{document}

\title{Dynamical scaling analysis of the optical Hall conductivity 
in the quantum Hall regime}

\author{Takahiro Morimoto}
\affiliation{Department of Physics, University of Tokyo, Hongo, 
Tokyo 113-0033, Japan}
\author{Yshai Avishai}
\affiliation{Department of Physics, Ben Gurion University, 
Beer Sheva 84105, Israel}
\affiliation{Department of Mathematics, 
University of Tokyo, Komaba, Tokyo 153-8914, Japan}
\author{Hideo Aoki}
\affiliation{Department of Physics, University of Tokyo, Hongo, 
Tokyo 113-0033, Japan}

\date{\today}

\begin{abstract}
Dynamical scaling analysis is theoretically performed for 
the  ac (optical) 
Hall conductivity $\sigma_{xy}(\varepsilon_F,\omega)$ 
as a function of Fermi energy $\varepsilon_F$ and frequency $\omega$
 for the two-dimensional electron gas 
and for graphene.  
In both systems, results based on exact diagonalization show that 
$\sigma_{xy}(\varepsilon_F,\omega)$ displays a well-defined 
dynamical scaling, for which the dynamical critical exponent 
as well as the localization exponent are fitted and plugged in. 
A crossover from the dc-like bahavior to the ac regime is  identified. 
The dynamical scaling analysis has enabled us to quantify 
the plateau in the ac Hall conductivity previously 
obtained, and to predict that the plateaux 
structure in ac is robust enough to be observed in the THz regime.  
\end{abstract}


\maketitle

{\it Introduction ---} 

Dynamics of electrons in the integer quantum Hall effect (QHE) 
is an interesting, hitherto not fully explored problem.  
Theoretically, the question is how the static 
Hall conductivity,  which may be regarded as a topological quantity\cite{tknn,hatsugai1993cna}, evolves into the optical Hall conductivity, especially in the THz regime where the relevant energy scale is the 
cyclotron energy.\cite{mikhailov85,morimoto-opthall,Fialkovsky09}  
Two of the present authors and Hatsugai have recently shown 
that the plateau structure in $\sigma_{xy}(\omega)$  is retained 
in the ac ($\sim$ THz) regime 
in both the ordinary two-dimensional electron gas (2DEG) and 
in graphene (described as the massless Dirac model), 
although the plateau height deviates from the quantized 
values in ac.\cite{morimoto-opthall}  
The numerical result indicates that the plateau structure remains remarkably robust 
against disorder, which can be attributed 
to an effect of localization which dominates the physics of electrons 
around the centers of Landau levels in disordered QHE systems.  
However, what is physically significant is not a result for a specific 
sample size, but the scaling behavior, especially when 
the localization is relevant in disordered systems.  For ac 
responses, we have to look into the dynamical scaling.   
Scaling analysis of  localization-delocalization transition 
 in the 2DEG QHE has been done for both 
 the static 
{\em longitudinal} conductivity 
  $\sigma_{xx}(\varepsilon_F)$, 
where $\varepsilon_F$ is the Fermi energy, \cite{huckestein} 
and for dynamical scaling properties of 
the longitudinal conductivity $\sigma_{xx}(\varepsilon_F,\omega)$,\cite{gammel-brenig}
but the dynamical scaling for the Hall conductivity $\sigma_{xy}(\varepsilon_F,\omega)$  has not been properly addressed 
for both ordinary and graphene QHE.   

With this motivation, here we elucidate the
dynamical scaling behavior of the ac Hall conductivity 
around the plateau to plateau 
transition to gain a deeper 
understanding of the optical Hall effect and its
robust step structures in the ac region.  
Namely, when we perform a scaling analysis for the plateau to plateau 
transition width $W$, the quantity depends on $\omega$.   
Physically, a new length scale, 
$L_\omega \sim \omega^{-\frac{1}{z}}$,  emerges  at finite 
frequencies, where $z$ is the 
dynamical critical exponent.  
We have performed the dynamical scaling analysis for 
both 2DEG and graphene QHE.  
The quantum Hall effect in graphene 
is unique in that a 
zero-energy 
Landau level (LL) exists, which has no counterpart in 
the QHE in 2DEG \cite{Nov05}. 
Thus the dynamic scaling is of special interest  
for the $n=0$ LL in graphene.   

Experimentally, scaling properties of $\sigma_{xx}(\varepsilon_F,\omega)$
was investigated from $\omega=0$ \cite{koch-klitzing91} 
up to the GHz regime \cite{hohls-kuchar02}.  
Recent advances 
in optical measurements (e.g., Faraday rotation in magnetic fields)  in the THz region have made 
the study of dynamical response functions feasible\cite{ikebe2008cds,ikebe-THz10}.  
For graphene, 
optical properties begin to be studied, among which 
are experimental transmission spectra\cite{sadowski}, or 
theoretical examination of the cyclotron emission\cite{morimoto-CE}.  
Thus, the physics of dynamical scaling in graphene QHE should 
be interesting in the THz regime.

Here we shall show that: 
(i) The ac Hall conductivity obeys a well-defined dynamical scaling.  
(ii) There is a crossover 
in the scaling behavior from a dc-like regime to an ac regime, 
in the latter of which $L_\omega $ dominates the scaling.   
In the former $L/\xi$ dominates the scaling (where $\xi$ is the localization length), while 
in the latter $L_\omega/\xi$ does.
(iii) The dynamical critical exponent is found to be $z \simeq 2$ 
in both the 2DEG and graphene QHE  systems as far as 
the potential disorder is concerned.  
(iv) The analysis enables us to estimate 
the plateau to plateau transition width $W$ 
in the ac regime with $L_\omega <L$ to assert that the Hall conductivity maintains the plateau structure at frequencies as high as $\omega\sim 0.1\omega_c$,  
which, for a magnetic field of a few Tesla, covers the 
 THz region.  This is an experimentally testable statement. 

{\it Formalism ---}  

For the ordinary QHE system as typically realized  in GaAs/AlGaAs,  
the kinetic part of the Hamiltonian is $H_0=\frac{1}{2 m^*}({\bf p} +e {\bf A})^2$,  
where  $m^*$ is the effective mass of the electron, ${\bf p}=(p_x,p_y)$ the momentum, and $ {\bf A}$ the vector potential. 
Disorder is introduced by a random potential $V({\bf r})$ composed of 
Gaussian scattering centers of range $d$ and density $n_{\mathrm{imp}}$ placed on randomly chosen points ${\bf R}_j$:
\begin{equation} \label{Didorder}
V({\bf r})=\sum_j u_j \exp(-|\Vec r- {\bf R}_j|^2/2d^2)/(2\pi d^2). 
\end{equation}
For $u_j$ we assumed a bimodal distribution $u_j=\pm u$ with  
random signs so that the broadened Landau level 
is symmetric.  A measure of disorder, i.e., the Landau 
level broadening\cite{ando}, is 
$
\Gamma = 2u [n_{\rm imp}/2\pi(\ell^2+d^2)]^{1/2}$.
Here we take $d=0.7\ell$, where $\ell=\sqrt{\hbar/eB}$ is the magnetic length,
but the result does not change significantly 
for other choices of $d$.
Diagonalization of the Hamiltonian is done for the subspace 
spanned by the five lowest LL's 
for $L\times L$ systems with $L/\ell$ varied over $25, 30, 35, 40$.
With wave functions and energy eigenvalues $\epsilon_a$ at hand, 
the optical Hall conductivity is evaluated from the Kubo formula, 
\begin{eqnarray}
\sigma_{xy}(\varepsilon_F,\omega)
&=&
\frac{i\hbar e^2}{ L^2}
\sum_{\epsilon_a < \varepsilon_F}\sum_{\epsilon_b \ge \varepsilon_F}
\frac 1 {\epsilon_b-\epsilon_a}
\nonumber\\
&\times&
\left(\frac{j_x^{ab}j_y^{ba}}{\epsilon_b-\epsilon_a-\hbar\omega} 
-\frac{j_y^{ab}j_x^{ba}}{\epsilon_b-\epsilon_a+\hbar\omega}
\right) ,
\label{kuboformula}
\end{eqnarray}
where $j_x^{ab}$ is the current matrix 
element\cite{morimoto-opthall}.  

The conductance is then averaged over 
a few thousands samples with different disorder potential realizations. 
The averaged conductivity  is hereafter denoted by the same symbol
 $\sigma_{xy}(\varepsilon_F,\omega)$.   For the scaling analysis 
the calculation done for varied sample size $L$, energy $\varepsilon_F$ and 
 frequency $\omega$.

For graphene QHE, we employ 
the two-dimensional effective Dirac model, 
\begin{equation} \label{HDirac}
H=v_F {\boldsymbol \sigma} \cdot {\boldsymbol \pi} + V({\bf r}),
\end{equation}
where ${\boldsymbol \sigma}=(\sigma_x,\sigma_y)$ is the 
Pauli matrices, ${\boldsymbol \pi}={\bf p}+e{\bf A}$, and
$V({\bf r})$ the random potential\cite{nomura2008qhe}.  
The selection rule for the current matrix elements in the Dirac model 
($|n| \leftrightarrow |n|\pm 1$ with $n$ the Landau index) is distinct
from that ($n \leftrightarrow n\pm 1$) for 2DEG.  

{\it Raw results for the Optical Hall Conductivity ---} 
\begin{figure}[tb]
\begin{center}
\begin{tabular}{cc}
\includegraphics[width=0.45\linewidth]{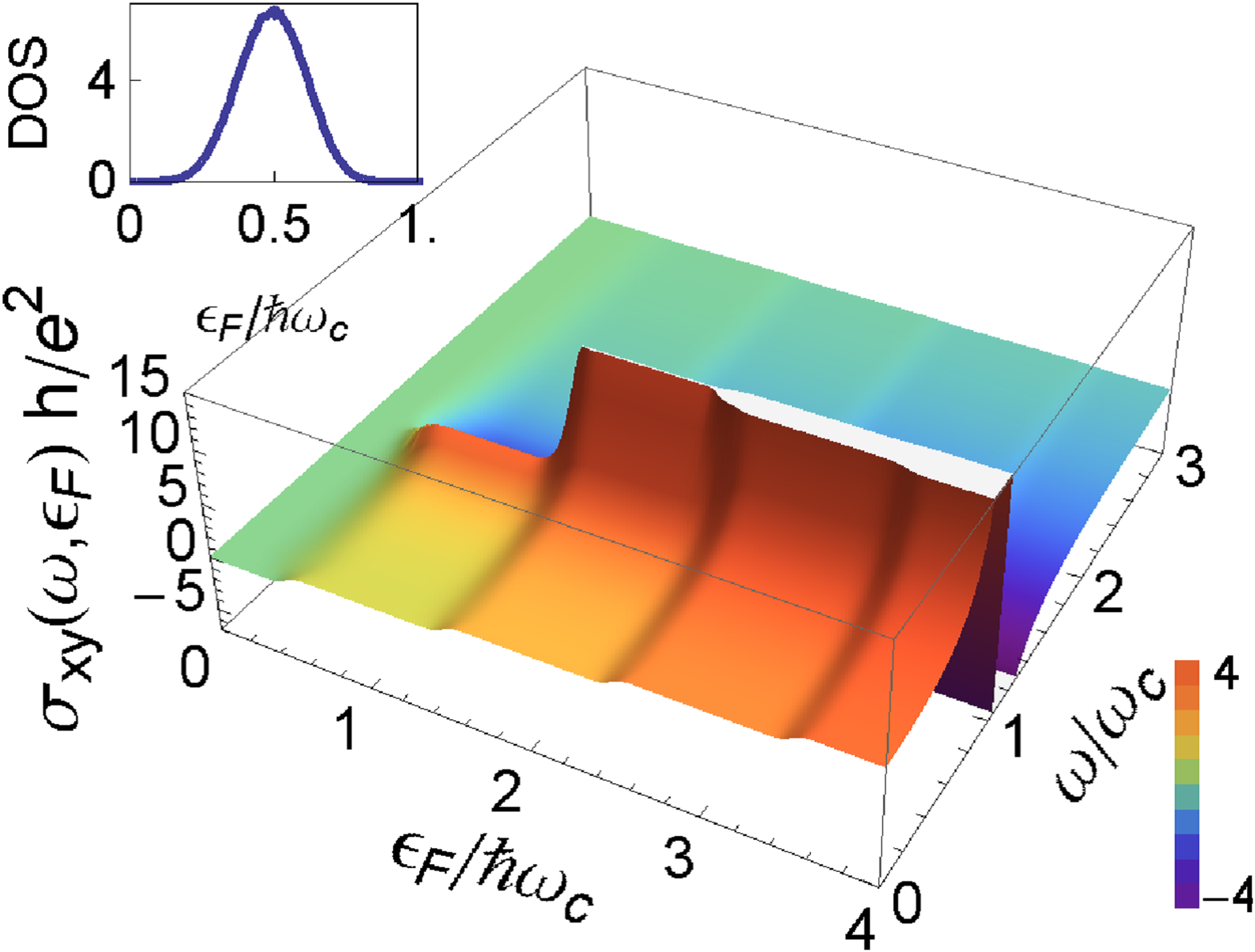}
&
\includegraphics[width=0.45\linewidth]{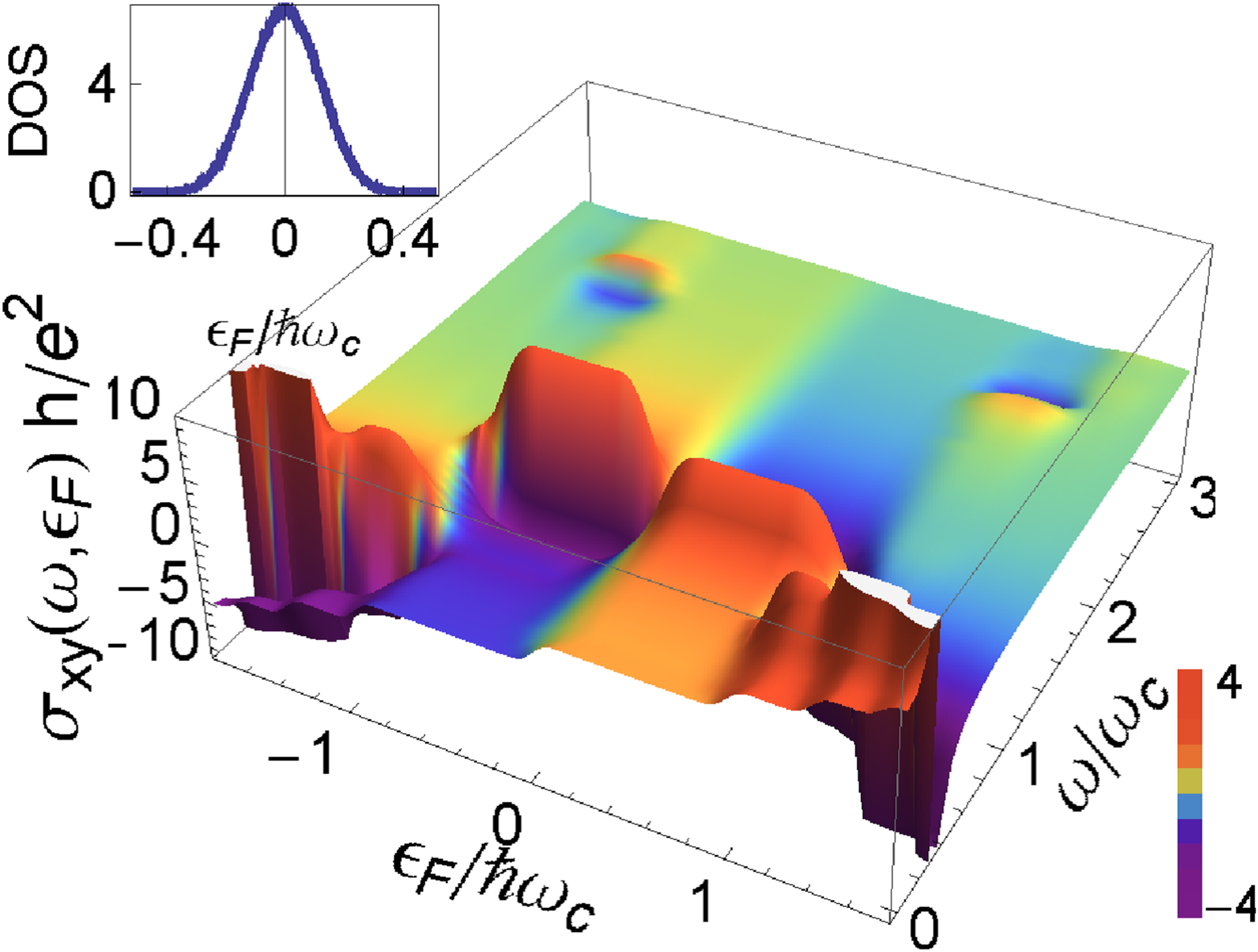}
\\
(a)&(b)\\

\includegraphics[width=0.45\linewidth]{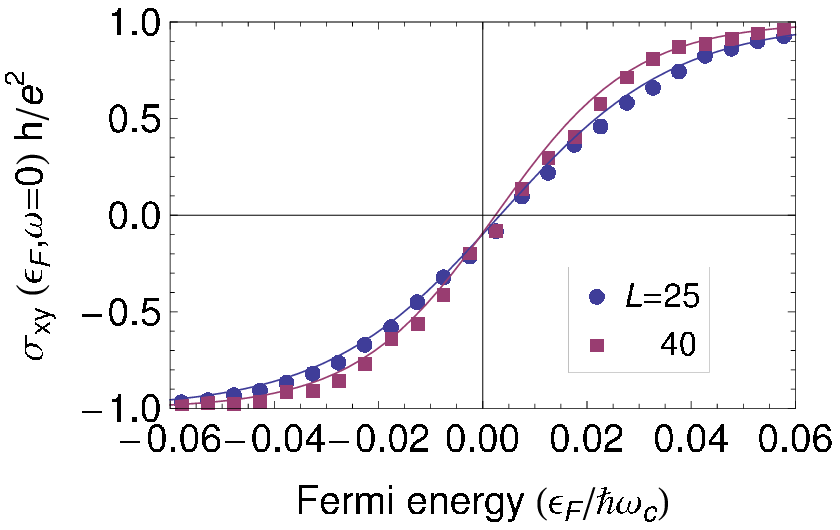}
&
\includegraphics[width=0.45\linewidth]{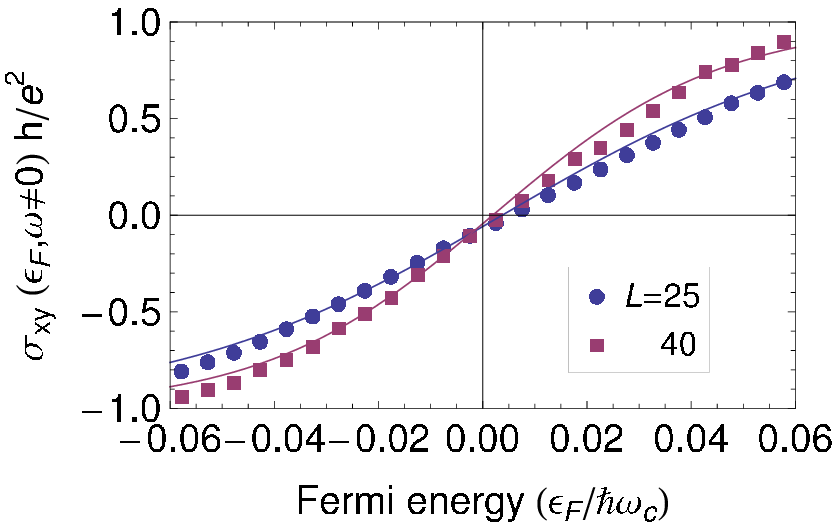}
\\
(c)&(d)\\
\end{tabular}

\end{center}

\caption{
$\sigma_{xy}(\varepsilon_F, \omega)$ plotted against Fermi energy $\varepsilon_F$ 
and frequency $\omega$ for (a) 2DEG, and (b) graphene quantum Hall systems for a pontential disorder with $\Gamma=0.4\hbar\omega_c$.
Insets are density of states for the lowest ($n=1$) LL (a)  and $n=0$ LL (b).
(c)The static Hall conductivity $\sigma_{xy}(\varepsilon_F)$ plotted against $\varepsilon_F$ for the graphene QHE system of sizes $L=25, 40$. (d) Optical Hall conductivity $\sigma_{xy}(\varepsilon_F,\omega)$ plotted against $\varepsilon_F$ for the graphene QHE system of sizes $L=25, 40$ 
for a fixed $\omega=6\omega_c/L^2$). The solid lines in (c) and (d) represent fitting with eqn.(\ref{tanh-fitting}).
}
\label{sxy3d}
\end{figure}

The optical Hall conductivity 
$\sigma_{xy}(\varepsilon_F,\omega)$ as a function of 
the Fermi energy $\varepsilon_F$ 
and frequency $\omega$ 
is displayed for the 2DEG (Fig.\ref{sxy3d}(a)) and 
graphene (Fig.\ref{sxy3d}(b)) QHE systems. 
The density of states (DOS) for Landau levels 
 (insets of Fig.\ref{sxy3d}(a,b)) confirms that 
the Landau level broadening 
is $\simeq \Gamma$. 
For each value of $\varepsilon_F$, the frequency-dependence 
is the Hall conductivity can be recognized as the cyclotron resonance.  
The 2DEG has one resonance 
at cyclotron frequency $\omega_c$ (Fig.\ref{sxy3d}(a)),
while the graphene QHE system exhibits a series of resonances, 
which correspond to 
the Dirac QHE selection rule $|n| \leftrightarrow |n|\pm 1$ (Fig.\ref{sxy3d}(b)) for the non-uniform set of Landau levels ($\propto \sqrt{n}$).  
Away from a resonance, {\it a step-like structure} is seen 
in $\sigma_{xy}(\varepsilon_F, \omega)$ as a function of $\varepsilon_F$ 
even for finite values of $\omega$, 
although the step heights are no longer quantized.  
We can attribute this behavior of $\sigma_{xy}(\varepsilon_F,\omega)$ 
to the localization property of electrons in QHE systems: As long as $\omega \ll \omega_c, \Gamma$, the nature of the mobility gap is maintained 
and $\sigma_{xy}$ remains flat.  
If we more closely insepct the width $W$ of the plateau to plateau 
transition, $W$ for a finite $\omega$  in Fig.\ref{sxy3d}(d) is 
seen to be greater than in the static case in 
Fig.\ref{sxy3d}(c), although still narrower than $\Gamma$.

We now move on to the sample size dependence 
of  the widths $W(\varepsilon,\omega,L)$ for the static and dynamic 
Hall conductivities.   
For the static Hall conductivity
$\sigma_{xy}(\varepsilon,0,L)$, 
we confirm the standard picture, where 
the plateau to plateau transition width becomes narrower
with the sample size $L$ as seen in Fig.\ref{sxy3d}.  
In the thermodynamic limit, almost all the wave functions are localized, where  the localization length 
diverges like $\xi \sim 1/|\varepsilon_F - \varepsilon_c|^\nu$ toward the center of the LL at $ \varepsilon= \varepsilon_c$.\cite{aoki-ando}  
For finite systems 
the states whose localization length $\xi$ is larger than the system size $L$ are effectively extended, and contribute to the longitudinal conductivity 
and the plateau to plateau transition.  This suggests 
the behavior $W \sim L^{-1/\nu}$.

{\it Scaling Analysis ---} 
We are now in position to look at the 
dynamical scaling analysis of the optical Hall conductivity $\sigma_{xy}(\varepsilon_F,\omega,L)$ and the width $W(\omega,L)$  of the plateau to plateau transition.  
We expect that the $W$ increasing with $\omega$ and decreasing 
with $L$ may be captured with some  scaling function.   
For that we have to quantify 
the width $W$, or the steepness ($\propto 1/W$) of the transition by
 fitting $\sigma_{xy}(\varepsilon_F,\omega,L)$ around the transition region
 for a given LL to some function of $\varepsilon_F$ for each value of $\omega$.   
To describe the transition $\sigma_{xy}/(-e^2/h) = 0 \rightarrow 1$ 
in the 2DEG QHE we take
\begin{eqnarray}
\sigma_{xy}(\varepsilon_F,\omega,L)_{\mathrm{2DEG}}=
\frac 1 2 + \frac 1 2 \tanh \left[\frac{\varepsilon_F-\frac 1 2 \hbar \omega_c}{W(\omega,L)} \right] ,
\label{artanQHE}
\end{eqnarray}
while for the transition $\sigma_{xy}/(-e^2/h) = -1 \rightarrow 1$ 
in graphene QHE we take    
\begin{eqnarray}
\sigma_{xy}(\varepsilon_F,\omega,L)_{\mathrm{graphene}}
=\tanh \left(\frac{\varepsilon_F}{W(\omega,L)} \right). 
 \label{tanh-fitting}
\end{eqnarray}
The quality of fitting of the plateau to plateau transition by the tanh function is quite satisfactory
\footnote{To be precise, we have included a slight shift of 
the center of the tanh function to have a better fit, but the shift is 
tiny $\sim 0.001 \hbar\omega_c$.}
 as can be seen in Fig.\ref{sxy3d}(c,d).

Dynamical scaling 
analysis for $\sigma_{xy}(\varepsilon_F,\omega,L)$ is carried out
in a similar manner as that for the longitudinal conductivity\cite{avishai-luck96}. 
In this ansatz the optical Hall conductivity is regarded to 
depend on Fermi energy and frequency 
only through the ratios $L/\xi$ and $L_\omega/\xi$.    
Here we have the localization length, $\xi \sim 1/|\varepsilon_F - \varepsilon_c|^\nu$ where $\varepsilon_c$ is the critical energy which 
coincides with the center of the LL, and 
 $L_\omega \sim 1/\omega^{1/z}$.  
Then the dynamical scaling ansatz for the 
optical Hall conductivity reads 
\begin{equation} \label{scalingsigmaxy}
\sigma_{xy}(\varepsilon_F,\omega,L)=\frac{e^2}{h} 
F((\varepsilon_F-\varepsilon_c) L^{1/\nu},\omega L^z),
\end{equation}
where $F$ is a universal scaling function.  
This implies that the width of the plateau to plateau transition 
scales as
\begin{equation}
W(\omega,L)=L^{-1/\nu}f(\omega L^z),
\label{scaling-eq}
\end{equation}
where $f$ is a universal function deduced from $F$.  
The first factor on the right-hand side makes 
the plateau to plateau transition width 
narrower for larger systems and dictates the dc scaling, 
while the second factor $f$ describes the dynamical scaling.

\begin{figure}[tb]
\begin{center}
\begin{tabular}{cc}
\includegraphics[width=0.5\linewidth]{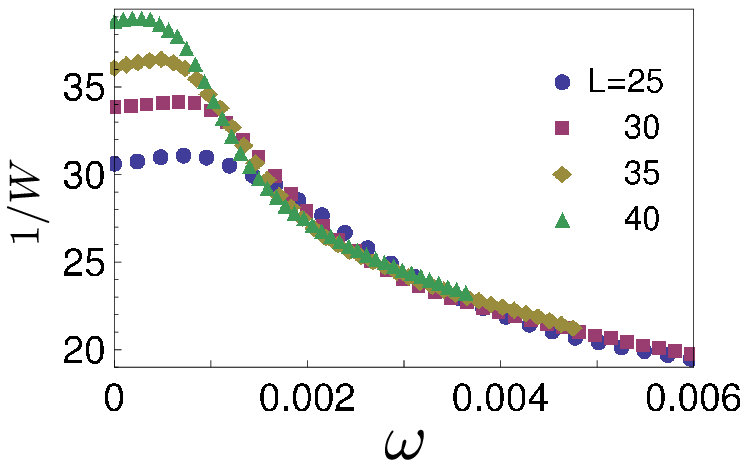}
&
\includegraphics[width=0.5\linewidth]{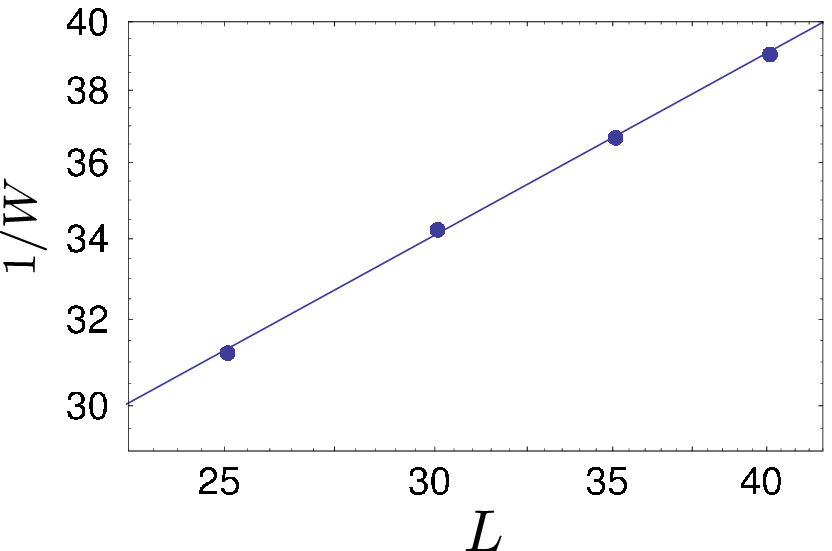}
\\
(a)&(b)
\\
\end{tabular}
\includegraphics[width=\linewidth]{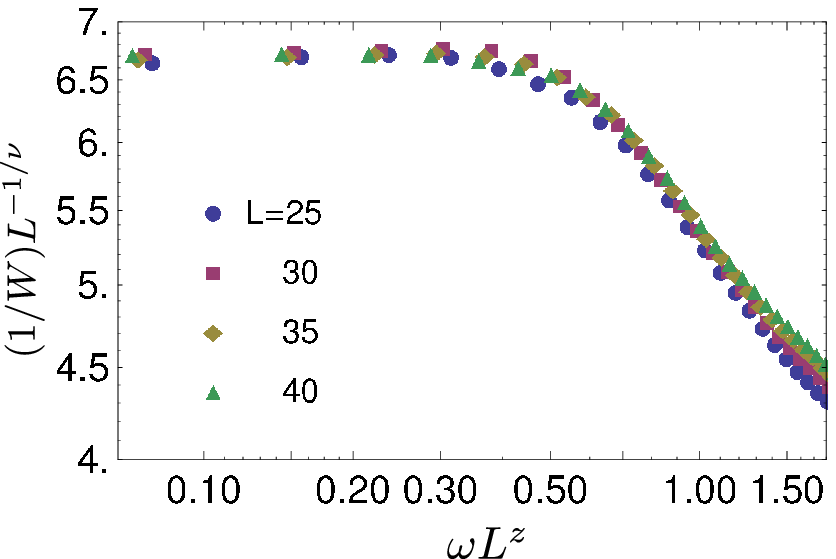}
\\
(c)
\end{center}

\caption{
Dynamical scaling analysis for the 2DEG QHE system
with $\Gamma=0.4\hbar\omega_c$ and $L= 25 - 40$: 
(a) The inverse width $1/W$ plotted against the frequency $\omega$, 
(b) the inverse width $1/W$ plotted against the sample 
size $L$, from which the localization exponent $\nu=2.1 \pm 0.2$ 
is obtained, and 
(c) rescaled inverse width $\frac{1}{W}L^{-1/\nu}$ plotted against the 
rescaled frequency $\omega L^z$ with a fitted dynamical 
critical exponent $z=1.8 \pm 0.2$.
$W,\omega,L$ are measured, respectively, 
in units of $\hbar\omega_c,\omega_c,\ell$. 
}
\label{ga-scaling}
\end{figure}
In Fig.\ref{ga-scaling} we show the scaling of inverse transition width, 
$1/W(\omega,L)$, for the 2DEG QHE system. 
By examining first the inverse width $ 1/W(\omega=0)$ for the static case against system size $ L$ in Fig.\ref{ga-scaling}(b), 
we obtain the localization critical exponent $\nu$ 
from $\log 1/W(\omega=0)=1/\nu \log L + f(0) $, with the result $\nu=2.1 \pm 0.2$.  This agrees with the accepted value of the static critical exponent in the integer QHE,  albeit slightly smaller.  

On the other hand, the frequency dependence of the inverse 
width for a fixed system size $L$ in (Fig.\ref{ga-scaling}(a))  
clearly exhibits that there are two regions: In the first region 
$1/W$ stays nearly constant up to some critical 
frequency that depends on the system size $L$, 
while in the other the quanitity begins to decrease monotonically with $\omega$.  
In the latter region, 
$1/W$ assumes similar values for all the sample sizes studied here 
as shown in (Fig.\ref{ga-scaling}(a)).   
We can indeed notice 
that, in the first region we have $L<L_\omega$, 
while in the second $L>L_\omega$.   
If we inspect 
 eqn.\ref{scaling-eq},  and assume a power-law form  
for the scaling function $f$, we can see that 
the inverse width in the second region should take a 
($L$-independent) form, 
$1/W(\omega) \propto \omega^{-1/z\nu}$. 
Calculation of  the dynamical exponent $z$ 
should be done for the 
critical region (i.e. for the transition width not too large), 
so that we do this around 
the crossing region where $1/W$ begins to 
decrease. This happens, 
typically, for $\omega<0.002\omega_c$.
With a least square fitting of $\log 1/W(\omega)= {\rm const} 
- \frac{1}{z\nu} \log \omega$, the 
dynamical critical exponent for the 2DEG QHE system 
is obtained as $z=1.8 \pm 0.2$ .
\begin{figure}[tb]
\begin{center}
\begin{tabular}{cc}
\includegraphics[width=0.5\linewidth]{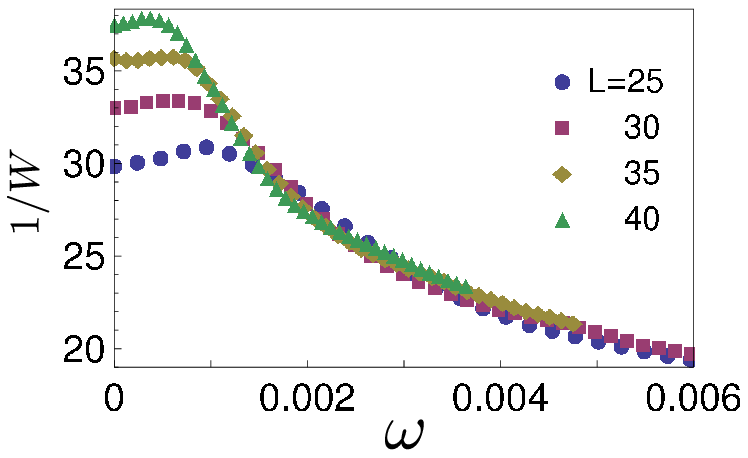}
&
\includegraphics[width=0.5\linewidth]{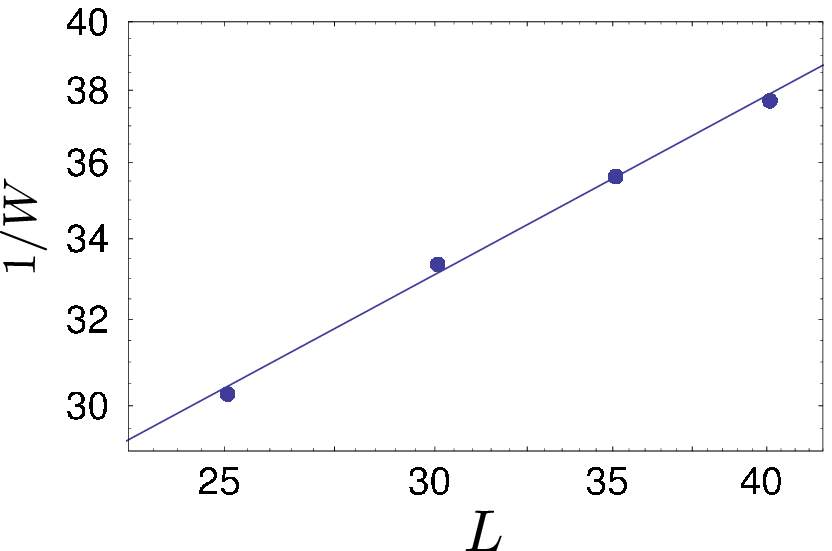}
\\
(a)&(b)
\\
\end{tabular}
\includegraphics[width=\linewidth]{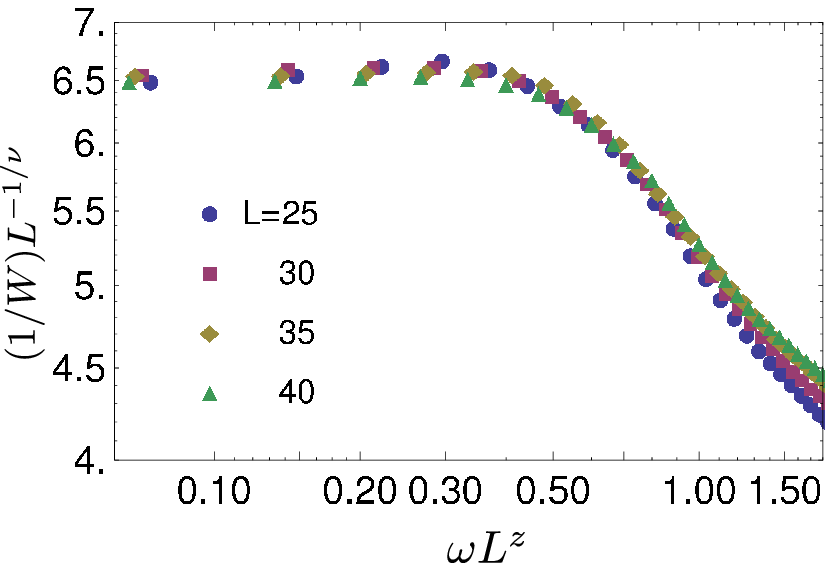}
\\
(c)
\end{center}

\caption{
Dynamical scaling analysis for the graphene QHE system
with $\Gamma=0.4\hbar\omega_c$ and $L= 25 
- 40$:  
(a) The inverse width $1/W$ plotted against the frequency $\omega$,
(b) the inverse width $1/W$ plotted against the sample 
size $L$, from which the localization exponent $\nu=2.1 \pm 0.1$ 
is obtained, and 
(c) rescaled inverse width $\frac{1}{W}L^{-1/\nu}$ plotted against the rescaled frequency $\omega L^z$ with a fitted dynamical 
critical exponent $z=1.8 \pm 0.2$.
}
\label{gr-scaling}
\end{figure}


If we now turn to the graphene QHE system, the same analysis 
is performed  based on Fig.\ref{gr-scaling}.  
The frequency dependence of the transition width is rather similar to 
that for the 2DEG system, as far as the potential disorder assumed 
here is concerned.  
 The localization exponent $\nu$ and and the dynamical critical exponent $z$ for the graphene system are determined as 
$\nu=2.1 \pm 0.1$  $z=1.8 \pm 0.2$, 
which coincide, within numerical errors, 
with those for the conventional QHE system.
This suggests that the two systems 
are in the same universality class.  
 As far as the dynamical exponent is concerned, it has been argued 
 by Hikami and Wegner \cite{Anderson-Localization-Solid-State-Sciences39}
  that, when the density of states is an analytic function of energy at a critical point, then $z=d$ ($d$: 
spatial dimension, which is 2 in the present case). Thus the fact that we get $z \approx 2$ for the 2DEG QHE is 
 commensurate with this conjecture. On the other hand, the density of states  for Dirac fermions ($\rho(E) \sim |E|$ for the clean system) 
 is non-analytic, for which one might expect 
a different behavior in graphene.   
The absence of deviation in the value of $z$ in Dirac fermions here 
should come from the fact that the presence of disorder smears the Dirac cone structure in the density of states to make it smooth (Fig.\ref{sxy3d}(b), inset). 


Having derived the static and dynamic exponents, 
we can now actually plot the 
scaling of the rescaled inverse width against rescaled frequency. 
This is displayed in Fig.\ref{ga-scaling}(c) for 2DEG, and Fig.\ref{gr-scaling}(c) (for graphene QHE).  
It can be judged that the scaling fit is quite good, which 
indicates that the form of 
the universal function assumed in eqn(\ref{scaling-eq}) is adequate.  
In the scaling plot we can see more clearly 
the first region with a constant $\frac{1}{W}L^{-1/\nu}$ for 
small $\omega L^z$, 
and the second region with a monotonously 
decreasing $1/W$ for larger $\omega L^z$. 

Intuitively, we can elaborate as follows:  
The dynamical response of the QHE system is governed by the 
magnitude of the localization length relative to two length scales: 
the system size $L$, and the length $L_\omega$ which 
is the distance over which an electron travels during one cycle, 
$1/\omega$, of the ac field.  
Since the localization length diverges as 
$\xi \sim |\varepsilon-\varepsilon_c|^\nu$  near the center of LL, 
and the states contributing to $\sigma_{xy}(\omega)$ should be those 
that simultaneously satisfy $\xi>L$ and $\xi>L_\omega$,
the transition width is determined by the smaller length scale $L$ or $L_\omega$.
In the static limit $\omega=0$, with $L_\omega \to \infty$, 
the system size $L$ determines the transition width $W$\cite{aoki-ando}.
When $\omega$ is increased, $L_\omega$ decreases.  
In the low enough frequency region, 
one still has $L_\omega > L$ so that the 
transition width continues to be  determined by the system size.  
When $L_\omega<L$ for higher frequencies, however, 
$W$ begins to be governed by $L_\omega$, 
and the transition width broadens monotonically with frequency.
For even higher frequencies, 
$L_\omega$ becomes so small that 
the system is far from the critical region,  and departure 
from the scaling  should occur.

From the functional form of $f(\omega L^z)$ in eqn(\ref{scaling-eq}),
the frequency for which $\omega L^z \sim 1$,
corresponds to the crossover region where the two regions overlap, 
 that is, $L_\omega \sim L$, 
or $L_\omega \sim 1/\omega^{1/z}$.
With the dynamical scaling argument with $z \simeq 2$, 
we end up with $L_\omega \sim 1/\omega^{1/2} \sim t^{1/2}$ where 
$t$ is the diffusion time. 
Since square-root time evolution is a characteristic of diffusion processes, 
the dynamical response behavior indicates that  the present  disordered system is diffusive.


The dynamical scaling here enables us to give an estimate of 
the transition width $W$ in the
THz region (with typically $\omega\sim 0.1\omega_c$).  
In the $L_\omega$-dominated regime, 
one obtains $W \propto \omega^{1/z\nu}$.  
The proportionality constant can be read out
from the numerical result, Fig. \ref{ga-scaling}(c),
so that
$W/\hbar\omega_c \sim 0.2 \omega^{1/z\nu} \sim 0.1$ at $\omega=0.1\omega_c$.  This implies that 
the plateau structure remains robust up to the THz region, so that  
experimental measurements should be feasible.

To summarize, dynamical scaling analysis 
of the optical Hall conductivity in 2DEG and graphene quantum Hall systems
have been carried out, 
while previous studies have focused mainly on the longitudinal conductivity.  
The dynamical critical exponent $z\simeq 2$ implies that 
the system is in the diffusive limit.  
The dynamical critical exponent $z$ is found to be similar 
between the 2DEG and the graphene QHE system, 
but we have to re-emphasize that this is as far as the pontential 
disorder taken here is concernded.  It is well-known\cite{kawarabayashi09}  that the preservation or otherwise of the chiral symmetry in 
disordered graphene has a profound effect on the $n=0$ 
graphene Landau level.  Thus it is an interesting future problem 
to look into this effect in terms of the dynamical scaling.


\end{document}